\documentstyle[11pt]{article}
\normalbaselines
\newcounter{eqnletter}[equation]
\catcode`\@=11
\@addtoreset{equation}{section}   % Makes \chapter reset 'equation' counter.
\catcode`\@=12

\begin{document}

\begin{center}

{\LARGE\bf Bethe Ansatz in Quantum Mechanics.}\\[0.7cm]
{\Large\bf 2. Construction of Multi-Parameter 
Spectral Equations}\footnote{This work was
partially supported by DFG grant no. 436 POL 113/77/0(S)}

\vskip 1cm

{\large {\bf Dieter Mayer} }

\vskip 0.1 cm

Institute of Theoretical Physics, TU Clausthal,\\
Arnold Sommerfeld Str. 6, 38-678 Clausthal, Germany\footnote{E-mail
address: dieter.mayer@tu-clausthal.de}\\

\vskip 1cm

{\large {\bf Alexander Ushveridze}  and {\bf Zbigniew Walczak}}

\vskip 0.1 cm

Department of Theoretical Physics, University of Lodz,\\
Pomorska 149/153, 90-236 Lodz, Poland\footnote{E-mail
addresses: alexush@mvii.uni.lodz.pl and walczak@mvii.uni.lodz.pl} \\

\end{center}
\vspace{1 cm}
\begin{abstract}

In this paper we propose a simple method for building
exactly solvable multi-parameter spectral equations
which in turn can be used for constructing completely
integrable and exactly solvable quantum systems.
The method is based on the use of 
a special functional relation which we call the scalar triangle equation 
because of its similarity to the classical Yang-Baxter
equation.

\end{abstract}
\newpage

\section{Introduction}

One of the possible ways for building completely integrable quantum systems 
is the {\it inverse method of separation of variables} based on  
the use of the 
so-called {\it multi-parameter spectral equations}. The idea is to 
interpret these equations as the result of separating variables in a 
certain multi-dimensional completely integrable quantum system and 
to reconstruct the form of the latter by eliminating some
of the spectral
parameters considered as separtion constants (for details
see e.g. refs. \cite{Us88,Us89,Us94, MaUsWa97}).

If one wants to obtain this way exactly 
solvable completely integrable systems,
one should start with {\it exactly solvable}
multi-parameter spectral equations. In this paper we present one possible 
way of building such equations by considering as an example  
one-dimensional second-order differential equations.
We show that in this case both the multi-parameter spectral equations 
and their solutions can be constructed from one and the same elementary 
building blocks called in the paper the {\it $\rho$-functions} which satisfy 
special functional relations which we called {\it scalar triangle equations}. 
The choice of this terminology is not accidental: The reader having some 
elementary acquaintance with the celebrated $r$-matrix approach 
(see e.g. \cite{Ji90}) can easily 
be convinced that there are many common features between the scalar triangle 
equations and ordinary matrix triangle equations\footnote{We mean here 
the so-called classical Yang -- Baxter equations.} There is also a deep 
relation between the solutions of these equations: the $\rho$-functions 
are natural analogues of the classical $r$-matrices. The general form of the 
substitution for 
solving the multi-parameter spectral equations is very similar 
to that of the famous Bethe Ansatz used for solving completely integrable 
quantum models. The numerical equations determining the solvability
conditions have the same meaning as the corresponding Bethe Ansatz equations.

In this paper we show that there are two essentially different classes of 
exactly solvable multi-parameter spectral equations which can be constructed 
from $\rho$-functions. We call them {\it rational} and {\it irrational} 
equations stressing the fact that in a certain "canonical" coordinate system 
their "potentials" can be expressed respectively via rational and irrational 
functions. The rational multi-parameter spectral equations are very well 
known in the literature. In 1987 Sklyanin obtained such equations 
\cite{Sk87} as a result
of a separation of variables in the so-called completely integrable Gaudin 
models \cite{Ga83} 
(which are easily constructable in the framework of the $r$-matrix 
approach). Applying to the rational multi-parameter spectral equations the 
inverse method of separation of variables one recovers the Gaudin models
as shown in refs. \cite{Us88,Us89,Us94}. As to the irrational multi-parameter 
spectral equations, these were never discussed in the literature. The reason
for this is that the form of the Bethe Ansatz equations 
determining their spectra drastically differs from the standard one usually 
obtained in the framework of $r$-matrix method. Moreover, there is no 
change of variables which could reduce these equations into standard form. 
This leads us to the claim that the class of completely integrable quantum 
systems associated with irrational multi-parameter spectral equations is
different from the known
classes of models and therefore its investigation is an 
interesting mathematical problem.

\section{The problem}

Second-order linear differential equations play an
important role in many branches of mathematical physics.
By an appropriate homogeneous transformation and a change
of variable any such equation can be reduced to the
following canonical form
\begin{eqnarray}
\left(-\frac{\partial^2}{\partial x^2}+W(x)\right)\psi(x)=0,
\label{1.1}
\end{eqnarray}
in which $W(x)$ and $\psi(x)$ are assumed to be analytic functions of 
the complex variable $x$. 

There are many different mathematical problems which are 
connected with an equation of the form (\ref{1.1}). 
The simplest one can be
formulated as follows: for a given function $W(x)$ find
the function $\psi(x)$. The general solution of this problem
obviously is
\begin{eqnarray}
\psi(x)\sim\sin\varphi \cdot\psi_1(x)+ \cos\varphi \cdot\psi_2(x)
\label{1.2}
\end{eqnarray}
where $\psi_1(x)$ and $\psi_2(x)$ are two linearly
independent solutions and $\varphi$ is an arbitrary mixing angle. 
The usual way to fix this angle is to impose one additional
constraint on the general solution (\ref{1.2}). This
constraint has to be compatible with the linearity of
equation (\ref{1.1}) and thus should have the form
\begin{eqnarray}
{\cal L}_0[\psi(x)]=0,
\label{1.3}
\end{eqnarray}
where ${\cal L}_0[\psi(x)]$ is some appropriately chosen
linear functional\footnote{In practice, one usually uses
the simplest functionals: ${\cal
L}_0[\psi(x)]\equiv \psi(0)$ and ${\cal L}_0[\psi(x)]\equiv \psi'(0)$.}.  

It is well known  that, along with this trivial (and
mathematically not very interesting) interpretation of equation (\ref{1.1}), 
there are many others which lead to richer sets of solutions and 
are of greater theoretical importance. The essence of most of these
interpretations is to allow some freedom in choosing
the function $W(x)$ restricting simultaneously the class of
allowed functions $\psi(x)$. This leads to the so-called
spectral versions of equation (\ref{1.1}).

Consider an example which will play a central role in our
further discussion. Thereby the form of the function $W(x)$ is
restricted to
\begin{eqnarray}
W(x)=W_0(x)+\sum_{n=1}^N e_n W_n(x),
\label{1.4}
\end{eqnarray}
where $W_0(x)$ and $W_1(x),\ldots,W_N(x)$ are some fixed
functions and $e_1,\ldots,e_N$ are arbitrary numerical parameters.
Restrict the class of admissible functions $\psi(x)$ by the following
$N+1$ constraints
\begin{eqnarray}
{\cal L}_0[\psi(x)]={\cal L}_1[\psi(x)]= \ldots
={\cal L}_N[\psi(x)]=0,
\label{1.5}
\end{eqnarray}
where the ${\cal L}_n[\psi(x)],\ n=0,\ldots, N$ are some arbitrarily chosen
linear but linearly independent functionals. Then one can state the problem 
of finding those values of the parameters $e_1,\ldots, e_N$ for
which equation (\ref{1.1}) (with $W(x)$ given by formula
(\ref{1.4})) has solutions fulfilling equations (\ref{1.5}).

It is natural to call $e_1,\ldots,e_N$ {\it spectral
parameters} and equation (\ref{1.1}) supplemented by conditions
(\ref{1.4}) and (\ref{1.5}) a {\it multi-parameter spectral equation}.
The set of admissible values for the parameters $e_1,\ldots,e_N$
we shall call the {\it spectrum}. It is easily seen that
the problem (\ref{1.1}), (\ref{1.4}), (\ref{1.5}) is a
natural generalization of an ordinary one-parameter spectral equation
which corresponds to the case $N=1$. 

It is not difficult to show that in general the spectrum of equations
(\ref{1.1}), (\ref{1.4}), (\ref{1.5}) is infinite and discrete. 
Indeed, let $\psi_1(x,e_1,\ldots,e_N)$ and $\psi_2(x,e_1,\ldots,e_N)$
denote two linearly independent solutions of equation (\ref{1.1})
considered as functions of the spectral parameters $e_1,\ldots,e_N$.
Then the general solution can be written in the form
\begin{eqnarray}
\psi(x)\sim \sin\varphi\cdot \psi_1(x,e_1,\ldots,e_N)+ \cos\varphi 
\cdot\psi_2(x,e_1,\ldots,e_N)
\label{1.6}
\end{eqnarray}
where $\varphi$ is an arbitrary parameter (mixing angle). 
Substituting (\ref{1.6}) into (\ref{1.5}) 
leads to a system of $N+1$ numerical equations
\begin{eqnarray}
l_0(\varphi,e_1,\ldots,e_N)=l_1(\varphi,e_1,\ldots,e_N)=
\ldots=l_N(\varphi,e_1,\ldots,e_N)=0,
\label{1.7}
\end{eqnarray}
for $N+1$ quantities $\varphi$ and $e_1,\ldots,e_N$. Here
$l_n(c,e_1,\ldots,e_N)$ denotes the value of the
linear functional ${\cal L}_n[\psi(x)]$ applied to the solution
(\ref{1.6}). Since the number of equations coincides with the 
number of unknowns, the spectrum of equation 
(\ref{1.1}), (\ref{1.4}), (\ref{1.5})
will be discrete in general. Also generally, the 
function (\ref{1.6}) is transcendental
which suggests that the spectrum should be infinite.

It is however quite obvious that the scheme given cannot be considered
a practical way for solving the multi-parameter spectral equations. 
This is so because for most functions $W_0(x)$ and $W_1(x),\ldots,W_N(x)$ 
the explicit form of the general solution (\ref{1.6}) is unknown.
Thus, one can expect that most of the multi-parameter spectral
equations of the form (\ref{1.1}), (\ref{1.4}), (\ref{1.5}) 
should not be exactly solvable.

Fortunately, there are exceptional cases, and these
we intend to discuss in this paper. We think of 
so-called {\it exactly solvable} multi-parameter spectral equations
which can be solved by means of purely algebraic methods
and have many important applications in the theory of
completely integrable quantum systems. One of the standard
and most effective methods for constructing such equations is
the so-called {\it inverse method}. It rests on a very
simple idea: instead of
looking for solutions of equation (\ref{1.1}) for
given functions $W_0(x)$ and $W_1(x),\ldots,W_N(x)$, one should 
try to reconstruct the form of these functions starting
with appropriately chosen function $\psi(x)$. An advantage
of the inverse problem in comparison with the direct one is
obvious: the problem of solving differential equations is
replaced by the problem of taking derivatives of known functions.
However, this is only a relative simplicity: the
algebro-analytic part of the inverse problem remains rather
non-trivial which is clearly seen from the following discussion:

Let us fix some $K+1$ functions $\psi_k(x),\ k=0,\ldots,K$
satisfying constraints (\ref{1.5}) and $K+1$ sets of numbers 
$\{e_{k1},\ldots,e_{kN}\},\ k=0,\ldots,K$.  
Substituting these into (\ref{1.1}), using relation (\ref{1.4}), 
taking $e_{k0}=1$ and dividing finally the 
$k$th equation by $\psi_k(x)$, we obtain
\begin{eqnarray}
\sum_{n=0}^N e_{kn} W_n(x)=\psi^{-1}_k(x)
\frac{\partial^2\psi_k(x)}{\partial x^2},\quad k=0,1,\ldots,K.
\label{1.8}
\end{eqnarray}
Formula (\ref{1.8}) can be considered a system of $K+1$ linear 
inhomogeneous equations for $N+1$ functions $W_0(x)$ and 
$W_1(x),\ldots,W_N(x)$. If $K\le N$, then system
(\ref{1.8}) is obviously solvable. In this case one easily finds 
the explicit form
of the functions $W_0(x)$ and $W_1(x),\ldots,W_N(x)$ for which
the equations (\ref{1.1}), (\ref{1.4}), (\ref{1.5}) have $K+1$
{\it a priori} known (and, hence, explicit) solutions. 
The situation changes, however, if $K>N$ (this is just our case, 
because we are looking for multi-parameter spectral equations having 
an infinite number of explicit solutions). In this
case, the number of equations (\ref{1.8}) exceeds the
number of unknowns which makes the system (\ref{1.8}) overdetermined
for almost all sets of functions $\psi_k(x)$. The only
way to get rid of this problem is to start with functions
$\psi_k(x)$ for which the compatibility 
of the equations forming system (\ref{1.8}) would be guaranteed 
from the very beginning. But for this to work 
we need some reasonable {\it ansatz} 
for the functions $\psi(x)$. It is hardly neccessary to emphasize
that the problem of finding such an ansatz is far from
being trivial.

From a purely practical point of view, it is much more
convenient to deal not with the functions $\psi(x)$ but with their 
logarithmic derivatives $P(x)$. In terms of the functions $P(x)$,
the ansatz which we intend to use takes an especially simple form. 
The substitution
\begin{eqnarray}
\psi(x)=\exp\left\{\int^x P(x')dx'\right\}
\label{1.9}
\end{eqnarray}
simplifies also the form of equation (\ref{1.1}). The
new equation
\begin{eqnarray}
W(x)=P^2(x)+P'(x),
\label{ste4.3}
\end{eqnarray}
will be considered as starting point for our further considerations.

\section{Separable functions of several variables}

We shall call a function $f(x_1,\ldots,x_k, y_1,\ldots,y_l)$ of $k+l$
variables {\it separable} with respect to the variables
$x_1,\ldots,x_k$ and $y_1,\ldots,y_l$ if
it can be represented in the form of a {\it finite} sum
\begin{eqnarray}
f(x_1,\ldots,x_k,y_1,\ldots,y_l)=\sum_{n=1}^Ng_n(x_1,\ldots,x_k)
h_n(y_1,\ldots,y_l)
\label{ste1.1}
\end{eqnarray}
where $h_n(x_1,\ldots,x_k),\ n=1,\ldots,N$ 
and $g_n(y_1,\ldots,y_l),\ n=1,\ldots,N$
are some functions of $k$ and $l$ variables, respectively. 
For stressing this property we shall denote such a function
by
\begin{eqnarray}
f(x_1,\ldots,x_k,y_1,\ldots,y_l)=f(x_1,\ldots,x_k|y_1,\ldots,y_l)
\label{ste1.2}
\end{eqnarray}
Functions of several variables separable with
respect to all arguments we shall call {\it
completely separable}. Any such function can be represented
in the form
\begin{eqnarray}
f(x_1,x_2,\ldots,x_m)=\sum_{n=1}^Nf_{1,n}(x_1)f_{2,n}(x_2)\ldots
f_{m,n}(x_m)
\label{ste1.3}
\end{eqnarray}
where the $f_{i,n}(x_i),\ i=1,\ldots,m$ are certain functions of one
variable. For such functions we shall use the notation
\begin{eqnarray}
f(x_1,x_2,\ldots,x_m)=f(x_1|x_2|\ldots|x_m)
\label{ste1.4}
\end{eqnarray}
Let us now formulate a simple lemma about separable
functions.

\medskip
{\bf Lemma 1.} Let $r(\xi)$ be a rational function of $\xi$.
Then 
\begin{eqnarray}
\frac{r(\xi_1)}{\xi_1-\xi_2}+
\frac{r(\xi_2)}{\xi_2-\xi_1}=f(\xi_1|\xi_2)
\label{ste1.5}
\end{eqnarray}
and
\begin{eqnarray}
\frac{r(\xi_1)}{(\xi_1-\xi_2)(\xi_1-\xi_3)}+
\frac{r(\xi_2)}{(\xi_2-\xi_3)(\xi_2-\xi_1)}+
\frac{r(\xi_3)}{(\xi_3-\xi_1)(\xi_3-\xi_2)}
=f(\xi_1|\xi_2|\xi_3)
\label{ste1.6}
\end{eqnarray}
are symmetric and completely separable functions.

\medskip
{\bf Proof.} Any rational function can be represented as a
linear combination of the so-called {\it elementary
rational functions} having only one singularity in the complex plane.
Thus, in order to prove separability of functions (\ref{ste1.5})
and (\ref{ste1.6}), 
it is sufficient to make sure that the positions of
the singularities of the functions $f(\xi_1|\xi_2)$ and 
$f(\xi_1|\xi_2|\xi_3)$ with respect to each 
argument do not depend on the values of other
arguments. It is clear that the "dangerous" singularities of both
functions (\ref{ste1.5}) and (\ref{ste1.6}) 
can only arise from their denominators. However, a
simple analysis shows that these singularities (which,
obviously, are present in each of the separate terms)
cancel each other. The absence of these
``dangerous'' singularities completes the proof of the lemma.

\section{Scalar triangle equation}

Let $r(\xi)$ be an arbitrary rational function of a
complex variable $\xi$. Define a complex valued function
$\xi(x)$ by the equation
\begin{eqnarray}
(\xi'(x))^2=r(\xi(x))
\label{ste2.1}
\end{eqnarray}
and construct from it a new function of two complex variables
\begin{eqnarray}
\rho(x,y)=\frac{1}{2}\cdot\frac{\xi'(x)+\xi'(y)}{\xi(x)-\xi(y)}.
\label{ste2.2}
\end{eqnarray}

\medskip
{\bf Lemma 2.} The function $\rho(x,y)$ obeys the
following functional relations:
\begin{eqnarray}
\rho(x,y)+\rho(y,x)=0,
\label{ste2.3}
\end{eqnarray}
\begin{eqnarray}
\rho(x,y)\rho(x,z)+\rho(y,z)\rho(y,x)+\rho(z,x)\rho(z,y)=\omega(x|y|z),
\label{ste2.4}
\end{eqnarray}
\begin{eqnarray}
\frac{\partial}{\partial x}\rho(x,y)+\rho^2(x,y)=\omega(x|x|y),
\label{ste2.5}
\end{eqnarray}
with $\omega(x|y|z)$ a certain separable function.

\medskip
{\bf Proof.} The proof of the anti-symmetry of the function $\rho(x,y)$
immediately follows from definition (\ref{ste2.2}). In order to prove
the separability of the function $\omega(x|y|z)$ we consider the following 
chain of equalities
\begin{eqnarray}
4\omega(x|y|z)=4[\rho(x,y)\rho(x,z)+\rho(y,z)\rho(y,x)+\rho(z,x)\rho(z,y)]=
\nonumber\\[0.6cm]
\frac{\xi'(x)+\xi'(y)}{\xi(x)-\xi(y)}\cdot
\frac{\xi'(x)+\xi'(z)}{\xi(x)-\xi(z)}+
\frac{\xi'(y)+\xi'(z)}{\xi(y)-\xi(z)}\cdot
\frac{\xi'(y)+\xi'(x)}{\xi(y)-\xi(x)}+
\frac{\xi'(z)+\xi'(x)}{\xi(z)-\xi(x)}\cdot
\frac{\xi'(z)+\xi'(y)}{\xi(z)-\xi(y)}=\nonumber\\[0.3cm]
\frac{[\xi'(x)]^2}{(\xi(x)-\xi(y))
(\xi(x)-\xi(z))}
+\frac{[\xi'(y)]^2}{(\xi(y)-\xi(z))
(\xi(x)-\xi(x))}
+\frac{[\xi'(z)]^2}{(\xi(z)-\xi(x))
(\xi(x)-\xi(y))}=\nonumber\\[0.3cm]
\frac{r[\xi(x)]}{(\xi(x)-\xi(y))(\xi(x)-\xi(z))}
+\frac{r[\xi(y)]}{(\xi(y)-\xi(z))(\xi(y)-\xi(x))}
+\frac{r[\xi(z)]}{(\xi(z)-\xi(x))(\xi(z)-\xi(y))}.\quad
\label{ste2.6}
\end{eqnarray}
Lemma 1 then shows that the function $\omega(x|y|z)$
is separable.
The last equation (\ref{ste2.5}) 
can be easily proved if we take $z=x+\epsilon$
in (\ref{ste2.4}), and take the limit
$\epsilon\rightarrow0$ noting that 
$\lim_{\epsilon\rightarrow0}\epsilon\rho(x+\epsilon,x)=1$. This
completes the proof.

\medskip
Hereafter we shall call equation (\ref{ste2.4}) the
scalar triangle equation.

\section{$\xi$-functions}

Any rational function of $\xi(x)$ and $\xi'(x)$ we
shall call a $\xi$-function. If $F(x)$ is a $\xi$-function,
it can be represented in the form
\begin{eqnarray}
F(x)=R(\xi(x))+\xi'(x)G(\xi(x))
\label{ste3.1}
\end{eqnarray}
where $R(\xi)$ and $G(\xi)$ are some rational functions.
The sum, difference, product and quotient of two
$\xi$-functions is again a $\xi$-function. The
derivative of a $\xi$-function is again a $\xi$-function. We call
a $\xi$-function $F(x)$ even if $G(\xi)\equiv 0$ in (\ref{ste3.1})
and odd if $R(\xi)\equiv 0$ in (\ref{ste3.1}). The product of two
odd or two even $\xi$-functions is an even $\xi$-function, and
the product of an even and an odd $\xi$-function is an odd $\xi$-function.
This means that the algebra of odd and even $\xi$-functions is
a $z_2$-graded algebra. The differential operator
$\partial/\partial x$ becomes in this case an odd object.

\medskip
{\bf Lemma 3.} If $F(x)$ is a $\xi$-function, then
\begin{eqnarray}
\sigma(x|y)=[F(x)-F(y)]\rho(x,y)
\label{ste3.2}
\end{eqnarray}
is a separable function.

\medskip
{\bf Proof.} Substituting (\ref{ste3.1}) into
(\ref{ste3.2}), we can write
\begin{eqnarray}
\sigma(x|y)=\sigma_R(x|y)+\sigma_G(x|y)
\label{ste3.3}
\end{eqnarray}
where
\begin{eqnarray}
\sigma_R(x|y)=\frac{R(\xi(x))-R(\xi(y))}{\xi(x)-\xi(y)}\cdot
(\xi'(x)-\xi'(y))
\label{ste3.4}
\end{eqnarray}
and
\begin{eqnarray}
\sigma_G(x|y)=\frac{\xi'(x)G(\xi(x))-\xi'(y)G(\xi(y))}
{\xi(x)-\xi(y)}\cdot (\xi'(x)-\xi'(y)).
\label{ste3.5}
\end{eqnarray}
From Lemma 1 it immediately follows that $\sigma_R(x|y)$ is
a separable function. In order to prove separability of
$\sigma_G(x|y)$, let us rewrite (\ref{ste3.5}) in the form
\begin{eqnarray}
\sigma_G(x|y)=\frac{[\xi'(x)]^2-[\xi'(y)]^2}
{\xi(x)-\xi(y)}\cdot (G(\xi(x))+G(\xi(y)))+
\frac{G(\xi(x))-G(\xi(y))}
{\xi(x)-\xi(y)}\cdot (\xi'(x)+\xi'(y))^2.
\label{ste3.6}
\end{eqnarray}
Using now (\ref{ste2.1}) in the first term and applying then
Lemma 1 we find that also $\sigma_G(x|y)$ is
separable. This proves the lemma.

\section{Bethe ansatz}

Let us look for solutions of equation (\ref{ste4.3}) in the
form
\begin{eqnarray}
P(x)=F(x)+\sum_{i=1}^M\rho(x,x_i)
\label{ste4.4}
\end{eqnarray}
where $M$ is an arbitrary non-negative integer, 
$x_1,\ldots,x_M$ are still unknown numbers and $F(x)$
is a $\xi$-function. We call
this form the Bethe Ansatz. 
Substituting (\ref{ste4.4}) into (\ref{ste4.3}) gives
\begin{eqnarray}
W(x)=F^2(x)+F'(x)+2\sum_{i=1}^M[F(x)-F(x_i)]
\rho(x,x_i)+\nonumber\\
+2\sum_{i=1}^MF(x_i)\rho(x,x_i)
+\sum_{i=1}^M\rho^2(x,x_i)+\sum_{i=1}^M\rho'(x,y)+
\sum_{i\neq k}^M \rho(x,x_i)\rho(x,x_k).
\label{ste4.5}
\end{eqnarray}
Using formulas (\ref{ste2.4}), (\ref{ste2.5}) and (\ref{ste3.2}),
we obtain
\begin{eqnarray}
W(x)=F^2(x)+F'(x)+2\sum_{i=1}^M\sigma(x|x_i)
+\sum_{i=1}^M\omega(x|x|x_i)+\sum_{i\neq k}^M \omega(x|x_i|x_k)+
\nonumber\\
+2\sum_{i=1}^M\rho(x,x_i)\left\{\sum_{k=1,k\neq i}^M \rho(x_i,x_k)
+F(x_i)\right\}.
\label{ste4.6}
\end{eqnarray}
We see that the first three terms in the right hand side of
(\ref{ste4.6}) represent some separable function of $x$ and
$x_i$, while the last sum of the so-called {\it unwanted terms} 
is, obviously, non-separable. In order to make
the function $W(x)$ separable, one should require 
all the unwanted terms to vanish. This is equivalent to the system
of $M$ equations
\begin{eqnarray}
\sum_{k=1,k\neq i}^M \rho(x_i,x_k)
+F(x_i)=0,\quad i=1,\ldots,M
\label{ste4.7}
\end{eqnarray}
which we shall call the Bethe Ansatz equations. If these
equations are satisfied then
\begin{eqnarray}
W(x)=F^2(x)+F'(x)+2\sum_{i=1}^M\sigma(x|x_i)
+\sum_{i=1}^M\omega(x|x|x_i)+\sum_{i\neq k}^M \omega(x|x_i|x_k)
\label{ste4.8}
\end{eqnarray}
and hence can be written in the form
\begin{eqnarray}
W(x)=W_0(x)+\sum_{n=1}^N e_nW_n(x)
\label{ste4.9}
\end{eqnarray}
where $W_0(x)$ and $W_n(x),\ n=1,\ldots,N$ 
are certain $\xi$-functions
and $e_n, \ n=1,\ldots, N$ are some numerical coefficients
in which all the dependence on numbers $x_1,\ldots,x_N$ is contained.
We see that equation (\ref{1.1}) takes in this case
the form of a multi-parameter spectral equation. The role
of the spectral parameters is played by the numbers $e_1,\ldots,e_N$.
The admissible values for these parameters are determined
by the solutions
of the Bethe Ansatz equations (\ref{ste4.7}). It is easily seen
that, for any finite $M$, the number of Bethe Ansatz
equations coincides with the number of unknowns and
therefore this system has a discrete set of solutions.
Since this is an algebraic system, it has a finite number of solutions
for any $M$, but $M$ was an arbitrary non-negative integer.
This means that the total number of solutions of system (\ref{ste4.7})
is infinite and thus the corresponding multi-parameter
spectral equation has infinite discrete and algebraically
calculable spectrum.

\section{Some simple examples}

In this section we consider three simple examples of
solutions of the scalar triangle equations and construct the
corresponding classes of $\xi$-functions.

\medskip
{\bf Example 1.} Assume that $r(\xi)$ is a first-order polynomial
\begin{eqnarray}
r(\xi)=a+b\xi.
\label{ste.1}
\end{eqnarray}
Then, from (\ref{ste2.1}) it follows that
\begin{eqnarray}
\xi'(x)=\sqrt{a+b\xi(x)}.
\label{ste.2}
\end{eqnarray}
Solving this differential equation we obtain
\begin{eqnarray}
\xi(x)=\frac{b(x-t)^2}{4}-\frac{a}{b}, \quad \xi'(x)=\frac{b(x-t)}{2}.
\label{ste.3}
\end{eqnarray}
Construction of the function $\rho(x,y)$ by formula (\ref{ste2.2})
gives
\begin{eqnarray}
\rho(x,y)=\frac{1}{x-y}.
\label{ste.4}
\end{eqnarray}
This function obeys all three relations (\ref{ste2.3}) -- (\ref{ste2.5})
with a trivial function $\omega(x|y|z)$:
\begin{eqnarray}
\omega(x|y|z)=0
\label{ste.5}
\end{eqnarray}
In this simple case, the set of $\xi$-functions coincides with
the set of all rational functions of $x$. 

\medskip
{\bf Example 2.} Let us now assume that $r(\xi)$ is a second-order
polynomial:
\begin{eqnarray}
r(\xi)=a+b\xi+c\xi^2.
\label{ste.6}
\end{eqnarray}
Then, from (\ref{ste2.1}) it follows that
\begin{eqnarray}
\xi'(x)=\sqrt{a+b\xi(x)+c\xi^2(x)}.
\label{ste.7}
\end{eqnarray}
Solving this differential equation we obtain
\begin{eqnarray}
\xi(x)=\frac{\sqrt{4ac-b^2}}{2c}\mbox{sinh}\
\sqrt{c}(x-t) -\frac{b}{2c},\quad
\xi'(x)=\frac{\sqrt{4ac-b^2}}{2\sqrt{c}}\mbox{cosh}\ \sqrt{c}(x-t).
\label{ste.8}
\end{eqnarray}
Construction of function $\rho(x,y)$ by formula (\ref{ste2.2})
gives in this case
\begin{eqnarray}
\rho(x,y)=\frac{\sqrt{c}}{2}\cdot\mbox{cth} \ \frac{\sqrt{c}}{2}(x-y).
\label{ste.9}
\end{eqnarray}
This function obeys all three relations (\ref{ste2.3}) -- (\ref{ste2.5})
with a constant function $\omega(x|y|z)$:
\begin{eqnarray}
\omega(x|y|z)=\frac{c}{4}.
\label{ste.10}
\end{eqnarray}
In this case, the set of $\xi$-functions coincides with
the set of all hyperbolic functions of $x$ with period
$2\pi i/\sqrt{c}$. For negative $c$ these functions become
trigonometric. 

\medskip
{\bf Example 3.} Let now $r(x)$ be a third-order polynomial of the form
\begin{eqnarray}
r(\xi)=a+b\xi+c\xi^2+d\xi^3
\label{ste.11}
\end{eqnarray}
Then, from (\ref{ste2.1}) it follows that
\begin{eqnarray}
\xi'(x)=\sqrt{a+b\xi(x)+c\xi^2(x)+d\xi^3(x)}.
\label{ste.12}
\end{eqnarray}
Solving this differential equation we obtain
\begin{eqnarray}
\xi(x)=\frac{4}{d}{\cal P}(x-t,g_2,g_3)-\frac{c}{3d}, \quad
\xi'(x)=\frac{4}{d}{\cal P}(x-t,g_2,g_3).
\label{ste.13}
\end{eqnarray}
where ${\cal P}(x,g_2,g_3)$ denotes the Weierstrass ${\cal P}$-function
with
\begin{eqnarray}
g_2=\frac{c^2-3bd}{12},\quad
g_3=\frac{2c^3+9bcd-27ad^2}{16\cdot 27}
\label{ste.14}
\end{eqnarray}
Construction of the function $\rho(x,y)$ by formula (\ref{ste2.2})
gives
\begin{eqnarray}
\rho(x,y)=\zeta(x-y,g_2,g_3)-\zeta(x-t,g_2,g_3)+\zeta(y-t,g_2,g_3),
\label{ste.15}
\end{eqnarray}
where $\zeta(x,g_2,g_3)$ denotes the Weierstrass $\zeta$-function.
This function obeys all three relations (\ref{ste2.3}) -- (\ref{ste2.5})
with a non-trivial function $\omega(x|y|z)$:
\begin{eqnarray}
\omega(x|y|z)={\cal P}(x,g_2,g_3)+{\cal P}(y,g_2,g_3)+{\cal P}(z,g_2,g_3)
\label{ste.16}
\end{eqnarray}
In this case, the set of $\xi$-functions coincides with
the set of all elliptic functions of $x$.

\section{An equivalent description}

Note that equation (\ref{ste4.3}) admits an
important equivalence transformation which preserves its
form and its spectrum. This transformation includes
the change of the initial variable $x$ 
\begin{eqnarray}
\xi=\xi(x)
\label{ste5.1}
\end{eqnarray}
and a linear inhomogeneous tranformation of the functions $P(x)$ and $W(x)$:
\begin{eqnarray}
\bar P(\xi)=\xi'(x)\left[P(x)+\frac{1}{2}\frac{\partial}{\partial
\xi(x)}\ln \xi'(x)\right],
\label{ste5.2}
\end{eqnarray}
\begin{eqnarray}
\bar W(\xi)=
\frac{W(x)}{[\xi'(x)]^2}-\frac{1}{2}\left(\frac{\partial}
{\partial \xi(x)}\right)^2 \ln \xi'(x) -
\frac{1}{4}\left(\frac{\partial}{\partial \xi(x)}\ln \xi'(x)\right)^2.
\label{ste5.3}
\end{eqnarray}
In terms of the new variable $\xi$ and new functions 
$\bar W$ and $\bar F$ the equation (\ref{ste4.3})
becomes indeed
\begin{eqnarray}
\bar W(\xi)={\bar P}^2(\xi)+{\bar P}'(\xi).
\label{ste5.4}
\end{eqnarray}
If we make in (\ref{ste5.4}) the substitution
\begin{eqnarray}
{\bar P}(\xi)=\frac{\bar\psi'(\xi)}{\bar\psi(\xi)}
\label{ste5.5}
\end{eqnarray}
then we obtain the transformed version of the initial linear
equation (\ref{1.1})
\begin{eqnarray}
\left(-\frac{\partial^2}{\partial \xi^2}+\bar W(\xi)\right)
\bar\psi(x)=0.
\label{ste5.6}
\end{eqnarray}
We see that the form of equation (\ref{ste5.6}) exactly
coincides with that of the initial equation (\ref{1.1}).
Linearity of the transformation (\ref{ste5.3}) means
that equation (\ref{ste5.6}) is again a multi-parameter spectral
equation having the same spectrum as (\ref{1.1}). 
The equations connected by the transformations (\ref{ste5.2}) and 
(\ref{ste5.3}) we shall call equivalent.

From the examples discussed in the previous section we know that 
the scalar triangle equation has many different solutions which lead
to exactly solvable multi-parameter spectral equations expressable
in terms of rational, trigonometric, elliptic and also more 
complicated functions. It is naturally to ask, which of these equations
are equivalent in the sense of the transformations (\ref{ste5.2}) --
(\ref{ste5.3}) and which are not. For this one should solve 
the classification problem. The simplest way to do this is to find
some distinguished variable $\xi=\xi(x)$ in terms of which the 
solutions of the triangle equation have a more or less unified form. The best
candidate for a variable is obviously the function
$\xi(x)$ obeying equation (\ref{ste2.1}). 
Indeed, in this case all functions $\rho(x,x_i)$ used in
the Bethe Ansatz (\ref{ste4.4}) transform
into functions with the same denominator
$\xi-\xi_i$, where $\xi=\xi(x)$ and $\xi_i=\xi(x_i)$. Furthermore,
the condition for vanishing of all unwanted (non-separable) terms in
expression (\ref{ste4.6}) transforms into the condition of 
regularity of this expression at the points $\xi=\xi_i$.
The most natural way to perform the neccessary calculations in this
case is to start out immediately from the transformed version (\ref{ste5.4}) 
of our equation and use only very general informations 
on the form of the functions $\bar W(\xi)$ and $\bar P(\xi)$. This
form follows from the results of section 5 and is given as
\begin{eqnarray}
\bar W(\xi)=A(\xi)+\sqrt{r(\xi)}B(\xi)
\label{ste5.7}
\end{eqnarray}
and
\begin{eqnarray}
\bar P(\xi)=a(\xi)+\sqrt{r(\xi)}b(\xi)
\label{ste5.8}
\end{eqnarray}
where $A(\xi)$, $B(\xi)$ and $a(\xi)$, $b(\xi)$ are some 
rational functions and $r(\xi)$ as in (\ref{ste2.1}). 
Substituting (\ref{ste5.7}) and (\ref{ste5.8}) into
(\ref{ste5.4}) we obtain two independent equations for $a(\xi)$
and $b(\xi)$
\begin{eqnarray}
A(\xi)=a^2(\xi)+a'(\xi)+r(\xi)b^2(\xi)
\label{ste5.9}
\end{eqnarray}
and
\begin{eqnarray}
B(\xi)=b'(\xi)+
\left(2a(\xi)+\frac{r'(\xi)}{2r(\xi)}\right)b(\xi)
\label{ste5.10}
\end{eqnarray}
In the following two sections we demonstrate that these
equations admit two principally different Bethe Ansatz
solutions which we call the {\it rational} and {\it
irrational} ones.

\section{The rational Bethe Ansatz solution}

Before discussing the general case, consider an important
special one. Obviously, the choice
\begin{eqnarray}
b(\xi)=0
\label{ste5.11}
\end{eqnarray}
leads to a simpler form for system (\ref{ste5.9}), (\ref{ste5.10}):
\begin{eqnarray}
A(\xi)=a^2(\xi)+a'(\xi),
\label{ste5.12}
\end{eqnarray}
and
\begin{eqnarray}
B(\xi)=0.
\label{ste5.13}
\end{eqnarray}
which is hence reduced to only one equation (\ref{ste5.12}). The Bethe
Ansatz for this equation reads
\begin{eqnarray}
a(\xi)=\alpha(\xi)+\sum_{i=1}^M \frac{1}{\xi-\xi_i}
\label{ste5.14}
\end{eqnarray}
Substituting this ansatz into (\ref{ste5.12}) and requiring
regularity of the function $A(\xi)$ at the points
$\xi_i,\ i=1,\ldots,M$ we obtain
\begin{eqnarray}
A(\xi)=\alpha^2(\xi)+\alpha'(\xi)+2\sum_{i=1}^M
\frac{\alpha(\xi)-\alpha(\xi_i)}{\xi-\xi_i}
\label{ste5.15}
\end{eqnarray}
where the numbers $\xi_i,\ i=1,\ldots,M$ have to
be solutions of the system of Bethe Ansatz equations
\begin{eqnarray}
\sum_{k=1,k\neq i}^M \frac{1}{\xi_i-\xi_k}+\alpha(\xi_i)=0,
\quad i=1,\ldots,M
\label{ste5.16}
\end{eqnarray}
Substituting (\ref{ste5.15}) and (\ref{ste5.13}) into (\ref{ste5.7})
and using Lemma 1 we can reduce the function $\bar W(\xi)$ to
the form
\begin{eqnarray}
\bar W(\xi)=\bar W_0(\xi)+\sum_{n=1}^N e_n \bar W_n(\xi)
\label{ste5.17}
\end{eqnarray}
with $\bar W_0(\xi)$ and $\bar W_n(\xi),\
n=1,\ldots,N$ certain rational functions and the numbers
$e_n,\ n=1,\ldots,N$ depending on the values of
parameters $\xi_i,\ i=1,\ldots,M$ which fulfill the Bethe
Ansatz equations (\ref{ste5.16}).

\section{The irrational Bethe Ansatz solution}

The most general form for the functions 
$a(x)$ and $b(x)$ for the Bethe Ansatz for $P(x)$ reads
\begin{eqnarray}
a(\xi)=\alpha(\xi)+\sum_{i=1}^M \frac{\alpha_i}{\xi-\xi_i}
\label{ste6.1}
\end{eqnarray}
and
\begin{eqnarray}
b(\xi)=\beta(\xi)+\sum_{i=1}^M \frac{\beta_i}{\xi-\xi_i}
\label{ste6.2}
\end{eqnarray}
where $\alpha_i,\beta_i,\xi_i,\ i=1,\ldots,M$ are some
unknown numerical parameters and $\alpha(\xi)$ and $\beta(\xi)$
are fixed rational functions. 
Substituting formulas (\ref{ste6.1}) and (\ref{ste6.2})
into equations (\ref{ste5.9}) and (\ref{ste5.10}) and
requiring regularity of the functions $A(\xi)$ and
$B(\xi)$ at the points $\xi_i,\ i=1,\ldots,M$ we obtain
\begin{eqnarray}
A(\xi)=\alpha^2(\xi)+\alpha'(\xi)+r(\xi)\beta^2(\xi)+
\sum_{i=1}^M
\frac{1}{(\xi-\xi_i)^2}\left(
\frac{r(\xi)-r(\xi_i)-(\xi-\xi_i)r'(\xi_i)}
{4r(\xi_i)}\right)+\nonumber\\
+\sum_{i=1}^M \frac{1}
{(\xi-\xi_i)}\left(
\alpha(\xi)-\alpha(\xi_i)+
\frac{r(\xi)\beta(\xi)-r(\xi_i)\beta(\xi_i)}{\sqrt{r(\xi_i)}}+
\frac{r(\xi)-r(\xi_i)}{4}\sum_{i=1}^M
\frac{[r(\xi_i)r(\xi_k)]^{-\frac{1}{2}}}{\xi_i-\xi_k}
\right)
\label{ste6.3}
\end{eqnarray}
and
\begin{eqnarray}
B(\xi)=\beta'(\xi)+\left(2\alpha(\xi)+
\frac{r'(\xi)}{4r(\xi)}\right)\beta(\xi)+\nonumber\\
+\sum_{i=1}^M \frac{1}{\xi-\xi_i}\left(
\beta(\xi)-\beta(\xi_i)+
\frac{\alpha(\xi)-\alpha(\xi_i)}{\sqrt{r(\xi_i)}}+
\frac{r'(\xi)/r(\xi)-r'(\xi_i)/r(\xi)}{4\sqrt{r(\xi_i)}}
\right)
\label{ste6.3'}
\end{eqnarray}
where the numbers $\xi_i,\ i=1,\ldots,M$ are again assumed to
be solutions of the system of Bethe Ansatz equations
\begin{eqnarray}
\sum_{k=1,k\neq i}^M \frac{1}{\xi_i-\xi_k}\left(
1+\sqrt{\frac{r(\xi_i)}{r(\xi_k)}}\right)+
\sqrt{r(\xi_i)}\beta(\xi_i)+\frac{r'(\xi_i)}{2r(\xi_i)}=0,
\quad i=1,\ldots,M
\label{ste6.4}
\end{eqnarray}
For the numbers $\alpha_i,\beta_i,\ i=1,\ldots,M$ we
obtain
\begin{eqnarray}
\alpha_i=\frac{1}{2},\quad
\beta_i=\frac{1}{2\sqrt{r(\xi)}}, \quad i=1,\ldots,M
\label{ste6.5}
\end{eqnarray}
Substituting (\ref{ste6.2}), (\ref{ste6.3}) and (\ref{ste6.3'})
into (\ref{ste5.7})
and using Lemma 1 we can reduce the function $\bar W(\xi)$ to
the form
\begin{eqnarray}
\bar W(\xi)=\bar W_0(\xi)+\sum_{n=1}^N e_n \bar W_n(\xi)
\label{ste6.6}
\end{eqnarray}
where $\bar W_0(\xi)$ and $\bar W_n(\xi),\
n=1,\ldots,N$ are certain irrational functions and the numbers
$e_n,\ n=1,\ldots,N$ depend on the 
parameters $\xi_i,\ i=1,\ldots,M$ which satisfy the Bethe
Ansatz equations (\ref{ste6.4}).

\end{document}